\documentclass[prb,twocolumn,amsmath,amssymb,floatfix,showpacs,superscriptaddress]{revtex4}

\usepackage{bm}

\usepackage{amssymb,amsmath}
\usepackage{graphics}
\usepackage{subfigure}
\usepackage{epstopdf}
\usepackage{epsfig}
\usepackage{color}
\usepackage{amsfonts}
\usepackage{bm}
\usepackage{amscd}
\usepackage{epsfig}
\usepackage{subfigure}
\usepackage{color}
\usepackage{color}
\definecolor{darkgray}{rgb}{0.3,0.3,0.3}
\definecolor{gray}{rgb}{0.5,0.5,0.5}
\definecolor{yellow}{rgb}{.4,.4,0}
\definecolor{orange}{rgb}{1,0.5,0}
\definecolor{darkgreen}{rgb}{0,0.5,0}
\definecolor{darkblue}{rgb}{0,0,1}
\definecolor{darkred}{rgb}{0.5,0,0}
\definecolor{purple}{rgb}{0.35,0,0.35}\newcommand{\up}{\uparrow}
\newcommand{\down}{\downarrow}

\def\sig{{\mbox{\boldmath{$\sigma$}}}}

\begin{document}
\title{
Mechanically-controlled spin-selective transport  }

\author{R. I. Shekhter}
\affiliation{Department of Physics, G\"{o}teborg University, SE-412 96 G\"{o}teborg, Sweden}

\author{ O. Entin-Wohlman}

\affiliation{Raymond and Beverly Sackler School of Physics and Astronomy, Tel Aviv University, Tel Aviv 69978, Israel}

\affiliation{Physics Department, Ben Gurion University,  Beer Sheva
84105, Israel}

\author{A. Aharony}


\affiliation{Raymond and Beverly Sackler School of Physics and Astronomy, Tel Aviv University, Tel Aviv 69978, Israel}
\affiliation{Physics Department, Ben Gurion University,  Beer Sheva
84105, Israel}\date{\today}

\begin{abstract}

A device enabling mechanically-controlled  spin and electric transport in  mesoscopic structures is proposed. 
It  is based on the transfer of electrons through weak links formed by suspended nanowires, on which the charge carriers experience a strong Rashba spin-orbit interaction that twists their spins. It is demonstrated that when the weak link bridges two magnetically-polarised electrodes, a significant spintro-voltaic effect takes place. Then,  by monitoring the generated voltage one  is able to measure  
electronic spins accumulated  in the electrodes, induced e.g., by circularly-polarised light, or alternatively, the amount of spin twisting. 
Mechanically-tuning  the device by  bending the nanowire  allows one to achieve full control over the spin orientations of the charge carriers. 

\end{abstract}


\maketitle

\section{Introduction}
\label{intro}

Achieving significant interplay among the electric, magnetic,  and mechanical degrees of freedom in solid-state devices suggests exciting perspectives in coherent operations involving all three of them. 
Charge carriers in conducting materials are perfect candidates for realising  such an interplay since they carry both electric charges and magnetic moments (their spins) and  can  be coupled quite strongly  to mechanical deformations in  beamlike mesoscopic setups. 

Indeed, experiments have demonstrated \cite{Sazonova,Lassagne,Steele,Chaste,Laird,Eichler}   the feasibility of   coupling charge carriers to mechanical vibrations of suspended nanodevices,  
showing, e.g., 
that a mechanically-vibrating single-walled carbon nanotube can also act concomitantly as a single-electron transistor. 
The role of the spin degree of freedom, 
 i.e.,  the generation, detection and exploitation of spin currents,  has been recently discussed quite extensively, in particular in conjunction with the spin Seebeck effect in the magnetic insulator
yttrium-iron garnet. \cite{SSE}
This effect refers to the generation of an electric power  from a temperature difference between 
the magnetic insulator and a layer of normal metal
attached to it.  \cite{Xiao,Adachi,Schreier,Cahaya} The temperature difference gives rise to a  spin current which is pumped into  the normal metal in a longitudinal  configuration, and  
induces there 
a  traverse  emf
via the inverse spin Hall effect. \cite{Hoffmann,Miao}  The companion phenomenon, i.e., the spin Peltier effect,  has been detected as well.  \cite{Flipse}
Thermally-activated spin current through ferromagnetic tunnel contacts has been detected in  Ref. \onlinecite{Dankert}.

A particularly-promising situation 
arises when the electric current through a mechanically-deformed weak link is provided by a battery of uncompensated electronic spins. Such a setup combines together all three types of degrees of freedom and allows for a plethora of intriguing phenomena.
When the magnetic polarisations in the electronic reservoirs forming the electrodes are not identical,  then quite generally  both charge and spin currents result from the  transport of electrons through the junction. 
The situation at hand  resembles in a way thermoelectric transport in a two-terminal junction: 
the two currents (charge and spin), flow in response to two affinities, the voltage difference and the difference in the amount of magnetic polarisation between  the two reservoirs.
``Non-diagonal"
phenomena, analogous to  the thermoelectric Seebeck and Peltier effects, can therefore be expected. For instance, it is possible to generate  a spin current by  injecting charges  into the material,  
which in turn may give rise to  a spatially inhomogeneous 
spin accumulation.

However, 
the two opposite spins can still contribute   equally   to the charge transport, 
resulting in zero net spin propagation, much like the vanishing of the thermopower when electron-hole symmetry is maintained. In the case of  combined spin and charge transport,   non-diagonal  spin-electric effects appear once the spin and charge transports are coupled in 
a way that distinguishes between the two spin projections.
One may achieve such a spin-dependent transport by exploiting  magnetic materials in which the electronic energy is spin-split. When the magnetization is spatially inhomogeneous (as happens in composite magnetic structures)  the spin-dependent part of the energy will be  inhomogeneous as well, leading to a spin-dependent force acting on the charge carriers. 
Another possibility, feasible   even in magnetically-homogeneous materials,  is to employ the     Rashba       spin-orbit  interaction \cite{Rashba} which can be controlled  by external electric fields. This interaction causes the electronic spin to rotate around an axis determined by its spin and  the electric-field direction. \cite{com}
When this interaction   varies in space, the electronic spin is twisted. 
The end result is the same as in the first scenario above: a spin-dependent force (resulting from the Rashba interaction) 
is exerted on the electrons,  opening the way for non-diagonal  spintro-electric transport.

Obviously, making such a spintro-electric effect tunable and controllable would be of great importance both from the viewpoint of fundamental physics as well as from that of practical applications. 
Here we propose that
such a manipulation of the spintro-electric transport can be achieved by
confining the spin-orbit interaction into a small  domain in space,  that at the same time can also be mechanically treated. 
In other words, one can modify geometrically the spatial region where  the spin-orbit interaction takes place.  When this very domain also serves as a weak link,   both the spin-orbit coupling and the electric resistance can be controlled through a geometrical deformation of the device.
Such an arrangement  can be realized in electronic weak links
or microconstrictions; it makes room for the possibility to control 
the transport by modifying the electronic  scattering 
in a small region of the material around the junction.
Specifically, we suggest that a suspended nanowire is most suitable for playing the role of the desired weak link. It is known that the Rashba spin-orbit interaction is anomalously large in certain nanowires \cite{nanowire} and also in nanotubes. \cite{SONT}
These beamlike 
structures are beneficial for our purposes since they are likely to produce  spin twist due to the Rashba interaction, while mechanically controlling their bending allows for the manipulation of 
the amount of twisting. This possibility arises because  mechanically bending  nanowires directly modifies the ballistic motion of the electrons through them, via the spin torque exerted  by the Rashba spin-orbit interaction. \cite{PRL}

Below we present a complete description of the spintro-electric transport through a Rashba spin-twister and demonstrate  the non-diagonal effects that are possible in such a device. 
Section \ref{EVT} presents the general formulation for the transport of the spin and the charge through a vibrating weak link, in the presence of both an Aharonov-Bohm flux and a Rashba spin-orbit interaction.  The results are summarized by the $3\times 3$ linear-response matrix of transport coefficients, Eqs. (\ref{22}) and (\ref{23}). Explicit expressions for these coefficients are derived in Sec. \ref{SV}, and given in Eq. (\ref{C1}). Section \ref{SV} also considers several special cases, showing how one can generate a voltage without a charge current  across an open circuit by a spin imbalance in the reservoirs (see Fig. \ref{1}), and how one can change the spin twisting by bending the weak link wire. Our conclusions are given in Sec. \ref{CON}. Certain detailed calculations are relegated to Appendices \ref{TUNHAM} and \ref{ME}.


\section{Spintro-voltaic effects due to Rashba splitting}


\label{EVT}
\subsection{General approach}
\label{GEN}

A ubiquitous description of transport phenomena through electronic weak links is based on the assumption that the electric resistance of the weak link dominates the resistance of the entire device. \cite{com3} This assumption  means 
that   the distribution of the electrons in momentum space in each of the electronic  reservoirs follows locally the equilibrium one. The electric current through the weak link is then accomplished by   tunnel coupling. Here we adopt this approach. 
However, having the electronic spin as an active component in the transport, this scheme   needs to be extended  to include also the distribution of the electrons in  spin space. The latter depends on the specific experimental setup.
For instance, injecting spin-polarised electrons  into each of the electrodes  when the spin-relaxation rate there is slow enough
yields ``spin pumping", \cite{Naaman} which results in an  imbalance between oppositely-oriented  electronic spins. Under these circumstances the electrochemical potential that determines the local equilibrium distribution in each of  the electrodes will be different for the two spin projections.
A similar situation can be created  upon using circularly-polarised light to pump excess spins into an electronic system. \cite{circ-pol} 
More options are open  when  the electrodes are made of magnetic materials. In that case the spin polarisation of the electrons  induced by the internal magnetization can differ from the one invoked by an external injection. The actual electronic distribution in spin space has then to be determined from an additional kinetic equation, a task which is beyond the scope of the present study. Instead, we will assume that the spin orientation of the injected electrons coincides with the direction of the internal magnetization in magnetic reservoirs. \cite{com1}


\begin{figure}[htp]
\includegraphics[width=8.5cm]{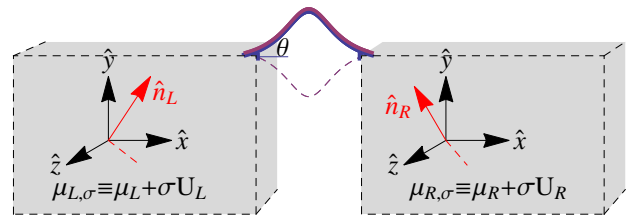}
\caption{ A schematic representation of the proposed setup. A nanowire, bended in the $x-y$ plane is coupled to two magnetically-polarised electronic reservoirs with arbitrarily-oriented magnetization axes $\hat{\bf n}_{L}$ and $\hat{\bf n}_{R}$.  The externally-pumped spins give rise to a spin-dependent electrochemical potentials $\mu_{L(R),\sigma}\equiv\mu_{L(R)} +\sigma U_{L(R)}$. The bending of the nanowire is specified by the angle it makes with the $\hat{\bf x}-$axis, with an instantaneous value $\theta$ around the equilibrium angle $\theta_{0}$.  } \label{1}
\end{figure}


The setup we propose is depicted schematically in Fig. \ref{1}. 
It comprises of
a nanowire 
bridging two leads, firmly coupled to the left and right electronic reservoirs,    held at spin-dependent electrochemical potentials $\mu_{L,\sigma}\equiv \mu_{L}+\sigma U^{}_{L}$ and $\mu_{R,\sigma}\equiv \mu_{R}+\sigma U_{R}$, respectively (the notations $L$ and $R$ refer to the left and the right leads, see Fig. \ref{1}).  Here $\sigma$ is the spin index.
The two bulk metals forming the reservoirs are each polarised along its own polarisation axis, denoted by the unit vectors  $\hat{\bf n}_{L}$ and $\hat{\bf n}_{R}$, respectively. The wire vibrates in the $x-y$ plane, such that the angle $\theta$ it makes with the $\hat{\bf x}-$axis oscillates around an equilibrium value,  $\theta_{0}$. An additional (weak) magnetic field,  applied along the $\hat{\bf z}-$direction, gives rise to an instantaneous Aharonov-Bohm effect, \cite{RIS} modifying  the transport properties  of the device and thus adding to its versatility.

The spin-resolved current through such a Rashba spin-junction was considered in detail in Ref. \onlinecite{PRL}. 
The model exploited in the explicit calculations
replaces the nanowire by a quantum dot \cite{wire} 
that has a single level, of energy $\epsilon_{0}$. 
As explained above, the reservoirs are represented by their respective electronic distributions  determined by the spin-dependent electrochemical potentials, 
\begin{align}
&f^{}_{L,\sigma}(\epsilon^{}_{k,\sigma})=[e^{\beta (\epsilon^{}_{k,\sigma}-\mu^{}_{L,\sigma})}+1]^{-1}\ ,\nonumber\\
&f^{}_{R,\sigma '}(\epsilon^{}_{p,\sigma '})=[e^{\beta (\epsilon^{}_{p,\sigma '}-\mu^{}_{R,\sigma '})}+1]^{-1}\ ,
\label{Fermi}
\end{align}
with $\beta^{-1}=k_{\rm B}T$.
The electron gas states
in the left (right) reservoir are indexed by $k,\sigma$ ($p,\sigma '$)  and have energies $\epsilon_{k,\sigma}$ ($\epsilon_{p,\sigma '}$).

\begin{figure}[htp]
\includegraphics[width=8cm]{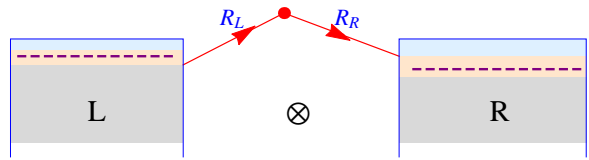}
\caption{Schematic geometry used for calculating the spin-orbit coupling dependence of the effective tunneling amplitude. A localized level 
is tunnel-coupled to left ($L$) and right ($R$) electronic electrodes. 
The setup lies in the $x-y$ plane; a magnetic field applied along $\hat{\bf z}$   is shown by $\otimes$. The radius vectors ${\bf R}_{ L}$ and ${\bf R}_{R}$ connect the quantum dot with the left and right electrodes.
}
\label{2}
\end{figure}

The linear Rashba interaction 
manifests itself by  phase factors multiplying  the tunneling amplitudes that couple the nanowire to the leads.\cite{SOhop}
In the geometry of Fig. \ref{2}, these phases are induced by an electric field perpendicular to the $x-y$ plane.  The phase factors are  given by
 $\exp[i\alpha{\bf R}\times\sig\cdot\hat{\bf z}]$, where     $\alpha$ denotes the strength of the spin-orbit interaction (in units of inverse length; units in which $\hbar=1$ are used),   and $\sig$ is the vector of the Pauli matrices.
Quite generally ${\bf R}={\bf R}_{L}\equiv \{x_{L},y_{L}\}$ for the left tunnel coupling  and   ${\bf R}={\bf R}_{R}\equiv \{x_{R},-y_{R}\}$
for the right one, where both radius vectors ${\bf R}_{L}$ and ${\bf R}_{R}$ are functions of the vibrational degrees of freedom (see Fig \ref{2}).
We adopt the plausible geometry  $y_{L}=y_{R}=(d/2)\sin(\theta)$ and $x_{L}=x_{R}=(d/2)\cos(\theta)$,  where $d$ is the wire length  ($\theta$ is the instantaneous bending angle).  \cite{footnote} In order to mimic the bending vibrations of the wire we assume that once the wire is bended by the  (equilibrium)  angle $\theta_{0}$,  then the distance along $x$ between the two supporting leads is fixed, while the (red) dot in Fig. \ref{2} vibrates along $y$.  As a result, ${\rm tan}(\theta) =2y/[d \cos(\theta)]$, implying that 
$\Delta\theta =(2/[d\cos (\theta)])\cos^{2}(\theta_{0})\Delta y$. [$d\cos (\theta^{}_{0})$ is the wire projection  on the $x-$direction.] 
It follows that 
\begin{align}
\theta=\theta_{0}+\Delta\theta=\theta_{0}+[a^{}_{0}\cos (\theta_{0})/d](b+b^{\dagger})\ , 
\label{theta}
\end{align}
where
$a_{0}$ is the amplitude of the zero-point oscillations, and
 $b$ ($b^{\dagger}$) is the destruction (creation) operator of the vibrations. Their free Hamiltonian is  described by the Einstein model, ${\cal H}_{\rm vib}=\omega b^{\dagger}b$.

The quantum vibrations of the wire, i.e. the dynamics of   the bending angle, 
make the electronic motion effectively two-dimensional.  \cite{RIS}   
This leads to the possibility of further  manipulating  the device   via the Aharonov-Bohm effect, by applying a magnetic field  perpendicular to the  junction plane (see Fig. \ref{2}). This field  imposes  an additional  phase on the tunneling
amplitudes $\phi_{L(R)}=-(\pi/\Phi_{0})(Hx_{L(R)}y_{L(R)})$ for the left and the right sides, respectively, where $H$ is the magnetic field 
(a factor of order unity  \cite{RIS}  is absorbed in $H$). The transport through the Rashba junction  depends only on the total Aharonov-Bohm phase, $\phi$, 
\begin{align}
\phi&\equiv\phi^{}_{L}+\phi^{}_{R}\nonumber\\
&=-\frac{\pi H}{\Phi^{}_{0}}(x^{}_{L}y^{}_{L}+x^{}_{R}y^{}_{R})=-\frac{\pi H d^{2}}{4\Phi^{}_{0}}\sin(2\theta )\ ,\label{phi}
\end{align}
measured in units of the flux quantum $\Phi_{0}=hc/e$.


\vspace{1.5cm}

\vspace{1cm}


The end result of the above considerations is that the tunneling through the Rashba weak link is effectively described by a tunneling Hamiltonian connecting directly the left and the right electrodes, \cite{PRL}
\begin{align}
{\cal H}^{\rm e}_{\rm tun}=
\sum_{k,p}\sum_{\sigma ,\sigma '}(c^{\dagger}_{p,\sigma '}[W^{}_{pk}]^{}_{\sigma '\sigma }c^{}_{k,\sigma }+{\rm H.c.})\ ,
\label{HTUN}
\end{align}
where $c^{}_{k,\sigma}$ and $c^{\dagger}_{k,\sigma}$  ($c^{}_{p,\sigma '}$ and $c^{\dagger}_{p,\sigma '}$)  are the annihilation and creation operators of the electrons  in the left (right) electrode. To second order in the (original) tunnelling amplitudes, the effective tunneling  is \cite{PRL}
\begin{align}
[W^{}_{pk}]^{}_{\sigma '\sigma }=\frac{1}{2}\sum_{\widetilde{\sigma}}V^{}_{p,\sigma '\widetilde{\sigma}}V^{}_{k,\widetilde{\sigma}\sigma}\Bigl (\frac{1}{\epsilon^{}_{k,\sigma}-\epsilon^{}_{0}}+\frac{1}{\epsilon^{}_{p,\sigma '}-\epsilon^{}_{0}}\Bigr )\ .\label{W}
\end{align}
The  tunneling amplitudes between the left and the right  electrodes and the quantum dot, $V_{p}$ and $V_{k}$ respectively, (matrices in spinor space)  consist of the ``bare" tunneling amplitudes (denoted $J_{L}$ and $J_{R}$, respectively), and the phases describing the effects of the perpendicular (Aharonov-Bohm) magnetic field and the Rashba interaction, 
\begin{align}
V^{}_{k(p)}=-J^{}_{L(R)}\exp[-i\psi^{}_{L(R)}]\ ,\label{vlr}
\end{align}
where
\begin{align}
\psi^{}_{L}&=\phi^{}_{L}-\alpha (x^{}_{L}\sigma^{}_{y}-y^{}_{L}\sigma^{}_{x})\ ,\nonumber\\
\psi^{}_{R}&=\phi^{}_{R}-\alpha (x^{}_{R}\sigma^{}_{y}+y^{}_{R}\sigma^{}_{x})\ .\label{psilr}
\end{align}
\begin{widetext}
The spin-resolved particle current emerging from the left electrode, $I_{L,\sigma}$, is  found \cite{PRL} by calculating the time evolution of the number operator of electrons with spin projection $\sigma$, $\dot{N}_{L,\sigma}$,
\begin{align}
\label{sol}
&- I^{}_{L,\sigma}\equiv\dot{N}^{}_{L,\sigma}=\int_{0}^{\infty} d\tau\sum_{k,p,\sigma '}
\Big [f^{}_{R,\sigma '}(\epsilon^{}_{p,\sigma '})[1-f^{}_{L,\sigma }(\epsilon^{}_{k,\sigma})]\\
&\times\Bigl (e^{i(\epsilon_{k,\sigma}-\epsilon_{p,\sigma '})\tau}\langle [W^{}_{pk}]^{}_{\sigma '\sigma}[W^{\dagger}_{kp} (\tau)]^{}_{\sigma \sigma '}\rangle +e^{i(\epsilon^{}_{p,\sigma '}-\epsilon^{}_{k,\sigma})\tau}
\langle [W^{}_{pk}(\tau)]^{}_{\sigma '\sigma}[W^{\dagger}_{kp}]^{}_{\sigma \sigma  '}\rangle\Bigr )\nonumber\\
&-f^{}_{L,\sigma}(\epsilon^{}_{k,\sigma})[1-f^{}_{R,\sigma '}(\epsilon^{}_{p,\sigma '})]\Bigl (e^{i(\epsilon^{}_{p,\sigma '}-\epsilon^{}_{k,\sigma})\tau}\langle [W^{\dagger}_{kp}]^{}_{\sigma\sigma '}[W^{}_{pk}(\tau)]^{}_{\sigma '\sigma}\rangle+e^{i(\epsilon^{}_{k,\sigma}-\epsilon^{}_{p, \sigma '})\tau}\langle [W^{\dagger}_{kp}(\tau )]^{}_{\sigma \sigma '}[W^{}_{pk}]^{}_{\sigma '\sigma}\rangle\Bigr )\Big ]\ .\nonumber
\end{align}
\end{widetext}
An analogous expression gives the current emerging from the right electrode. 
The angular brackets in Eq. (\ref{sol}) denote thermal averaging over the vibrations and over their time evolution with respect to the Einstein Hamiltonian. Assuming off-resonance conditions, the wave vector-dependence of the effective tunneling amplitude may be discarded, and then   [see Eq. (\ref{W})]
\begin{align}
&\langle [W^{}_{pk}]^{}_{\sigma '\sigma}[W^{\dagger}_{kp} (\tau)]^{}_{\sigma \sigma '}\rangle =\frac{J^{2}_{L}J^{2}_{R}}{\epsilon^{2}_{0}}\nonumber\\
&\times\langle[e^{-i\psi^{}_{R}}e^{-i\psi^{}_{L}}]^{}_{\sigma '\sigma}
[e^{i\psi^{\dagger}_{L}(\tau)}e^{i\psi^{\dagger}_{R}(\tau)}]^{}_{\sigma\sigma '}\rangle\ ,\label{ww}
\end{align}
[note that $\psi^{}_{L,R}=\psi^{\dagger}_{L,R}$, see Eqs. (\ref{psilr})] with
\begin{align}
&\langle[e^{-i\psi^{}_{R}}e^{-i\psi^{}_{L}}]^{}_{\sigma '\sigma}
[e^{i\psi^{\dagger}_{L}(\tau)}e^{i\psi^{\dagger}_{R}(\tau)}]^{}_{\sigma\sigma '}\rangle
\nonumber\\
&=\sum_{n,n'}P(n)
e^{i(n'-n)\omega\tau}
|\langle n|[e^{-i\psi^{}_{R}}e^{-i\psi^{}_{L}}]^{}_{\sigma '\sigma}|n'\rangle|^{2}\ .\label{tp}
\end{align}
Here $|n\rangle$ and $|n'\rangle$ denote the eigenstates  of energies $(n+1/2)\omega$ and $(n'+1/2)\omega$, respectively, of the Einstein vibrations, and
\begin{align}
P(n)=\frac{e^{-(n+1/2)\beta\omega }}{{\rm Tr} e^{-\beta {\cal H}^{}_{\rm vib}}}=e^{-n\beta\omega}(1-e^{-\beta\omega})\ ,\label{pn1}
\end{align}
such that
$
\sum_{n=0}^{\infty}P(n)=1$ and  $ \sum_{n=0}^{\infty}P(n)n=1/[e^{\beta\omega}-1]\equiv N^{}_{B}(\omega )$ is the Bose-Einstein distribution.
The other thermal averages in Eq. (\ref{sol}) are expressed in a similar form.

Since the electrodes are  magnetically-polarised, the density of states in each of them depends on both the internal exchange interaction and the external spin pumping as expressed by the energy split of the electrochemical potentials $U_{L,R}$ that determine the kinetic energy of the electrons participating in the transport. However, the latter dependence is 
weak,  \cite{com2m}      
and to lowest order in $U_{L,R}/\mu$, where $\mu=(\mu_{L}+\mu_{R})/2$ is the common chemical potential of the entire device, it may be neglected. Therefore one is able to convert  the sums over the wave vectors in Eq. (\ref{sol}) into integrals by introducing the spin-resolved densities of states at the common chemical potential of the left and right leads, ${\cal N}_{L,\sigma}$ and ${\cal N}_{R,\sigma}$, respectively.  It then  turns out that the spin-resolved particle currents emerging from the left and the right electrodes are
\begin{align}
&-I^{}_{L,\sigma}=2\pi{\cal N}^{}_{L,\sigma}\sum_{\sigma '}{\cal N}^{}_{R,\sigma '}\sum_{n,n'=0}^{\infty}
P(n){\cal T}^{}_{nn',\sigma\sigma '}\nonumber\\
&\times
(1-e^{\beta(\mu^{}_{L,\sigma}-\mu^{}_{R,\sigma '})})\frac{\mu^{}_{L,\sigma}-\mu^{}_{R,\sigma '}+(n'-n)\omega
}{e^{\beta[\mu_{L,\sigma}-\mu_{R,\sigma '}+(n'-n)\omega]}-1}\ ,\label{ils}
\end{align}
and 
\begin{align}
&-I^{}_{R,\sigma '}=2\pi{\cal N}^{}_{R,\sigma '}\sum_{\sigma }{\cal N}^{}_{L,\sigma }\sum_{n,n'=0}^{\infty}
P(n){\cal T}^{}_{nn',\sigma\sigma '}\nonumber\\
&\times
(e^{\beta(\mu^{}_{L,\sigma}-\mu^{}_{R,\sigma '})}-1)\frac{\mu^{}_{L,\sigma}-\mu^{}_{R,\sigma '}+(n'-n)\omega
}{e^{\beta[\mu_{L,\sigma}-\mu_{R,\sigma '}+(n'-n)\omega]}-1}\ .\label{irs}
\end{align}
Clearly  particle number is conserved, as can be seen by adding together Eq. (\ref{ils}) summed over $\sigma$ and Eq. (\ref{irs}) summed over $\sigma '$.


The spin indices of the matrix element squared \cite{com2}   forming the transmission,  ${\cal T}$, in Eqs. (\ref{ils}) and (\ref{irs}) deserve some caution:  the quantization axes of the magnetization in the two electronic reservoirs are generally different (see Fig. \ref{1}), and they both may differ from the quantization axis which is used to describe the Rashba interaction on the nanowire. Specifying the quantization axis in the left  (right) reservoir  by the angles $\theta_{L}$ ($\theta_{R}$) and $\varphi_{L}$ ($\varphi_{R}$), then
\begin{align}
{\cal T}^{}_{nn',\sigma\sigma '}=\Big (\frac{J^{}_{L}J^{}_{R}}{\epsilon^{}_{0}}\Big )^{2}_{}|\langle n|[{\cal S}^{\dagger}_{R}e^{-i\psi^{}_{R}}e^{-i\psi^{}_{L}}{\cal S}^{}_{L}]^{}_{\sigma '\sigma}|n'\rangle|^{2}
\ , \label{T}
\end{align}
where the rotation transformations ${\cal S}_{L(R)}$ are given by
\begin{align}
&{\cal S}^{}_{L(R)}=\nonumber\\
&\left [\begin{array}{cc}
e^{-i\frac{\varphi^{}_{L(R)}}{2}}\cos \frac{\theta^{}_{L(R)}}{2}
&
e^{-i\frac{\varphi^{}_{L(R)}}{2}}\sin \frac{\theta^{}_{L(R)}}{2}
\\
e^{i\frac{\varphi^{}_{L(R)}}{2}}\sin
 \frac{\theta^{}_{L(R)}}{2})
&
-e^{i\frac{\varphi^{}_{L(R)}}{2}}\cos \frac{\theta^{}_{L(R)}}{2}
\end{array}\right ]\ .\label{slr}
\end{align}
For instance, when the quantization axes in both electrodes are identical, $\hat{\bf n}_{L}=\hat{\bf n}_{R}$, 
then ${\cal S}$ just rotates the direction of the  quantization axis of the 
Rashba interaction.

\subsection{The linear-response regime}

\label{LR}

As is mentioned above, the transport of the charge carriers in our setup consists of both charge and spin currents. Here we examine these currents in 
the linear-response regime, where the spin-resolved particle currents, Eqs.  (\ref{ils}) and (\ref{irs})
become
\begin{align}
&I^{}_{L,\sigma}=2\pi{\cal N}^{}_{L,\sigma }\sum_{\sigma '}{\cal N}^{}_{R,\sigma '}(\mu^{}_{L,\sigma }-\mu^{}_{R,\sigma '}){\cal A}^{}_{\sigma\sigma'}\ ,\nonumber\\
&I^{}_{R,\sigma '}=2\pi{\cal N}^{}_{R,\sigma '}\sum_{\sigma }{\cal N}^{}_{L,\sigma }(\mu^{}_{R,\sigma ' }-\mu^{}_{L,\sigma }){\cal A}^{}_{\sigma\sigma'}\ ,
\label{lr}
\end{align}
with the transmission
\begin{align}
{\cal A}^{}_{\sigma\sigma '}&=\sum_{n=0}^{\infty}
P(n){\cal T}^{}_{nn,\sigma\sigma '} \nonumber\\
&+\sum_{\stackrel{nn'=0}{n\neq n'}}^{\infty}P(n){\cal T}^{}_{nn',\sigma\sigma '}
\frac{
(n'-n)\beta\omega
}{e^{(n'-n)\beta\omega}-1}\ .\label{a}
\end{align}
The first term in Eq. (\ref{a}) gives the contribution to the spin-resolved transport  from the elastic tunneling processes.
The second is due to the inelastic processes, and is active at finite temperatures.

Our final expressions for the charge currents are then
\begin{align}
eI^{}_{L}&\equiv e\sum_{\sigma}I^{}_{L,\sigma}=e(\mu^{}_{L}-\mu^{}_{R}){\cal C}_{1}-eU^{}_{R}{\cal C}^{}_{3}+eU^{}_{L}{\cal C}^{}_{2}\ ,
\label{II}
\end{align}
with $eI_{R}\equiv e\sum_{\sigma '}I_{R,\sigma '}=-eI_{L}$. The spin currents emerging from the left and right reservoirs are
\begin{align}
I_{L}^{\rm spin}\equiv \sum_{\sigma}\sigma I^{}_{L,\sigma}=
(\mu^{}_{L}-\mu^{}_{R}){\cal C}^{}_{2}-U^{}_{R}{\cal C}^{}_{4}+U^{}_{L}{\cal C}^{}_{1}\ ,\nonumber\\
I_{R}^{\rm spin}=\sum_{\sigma'}\sigma' I^{}_{R,\sigma' }=
(\mu^{}_{R}-\mu^{}_{L}){\cal C}^{}_{3}+U^{}_{R}{\cal C}^{}_{1}-U^{}_{L}{\cal C}^{}_{4}\ .\label{IIS}
\end{align}
In Eqs. (\ref{II}) and (\ref{IIS}) we have introduced the  linear-response transport coefficients
\begin{align}
{\cal C}^{}_{1}&=2\pi\sum_{\sigma\sigma '}{\cal N}^{}_{L,\sigma}{\cal A}^{}_{\sigma\sigma '}{\cal N}^{}_{R,\sigma '}\ ,\nonumber\\
{\cal C}^{}_{2}&=2\pi\sum_{\sigma\sigma '}{\cal N}^{}_{L,\sigma}\sigma{\cal A}^{}_{\sigma\sigma '}{\cal N}^{}_{R,\sigma '}\ ,
\nonumber\\
{\cal C}^{}_{3}&=2\pi\sum_{\sigma\sigma '}{\cal N}^{}_{L,\sigma}{\cal A}^{}_{\sigma\sigma '}\sigma '{\cal N}^{}_{R,\sigma '}\ ,\nonumber\\
{\cal C}^{}_{4}&=2\pi\sum_{\sigma\sigma '}{\cal N}^{}_{L,\sigma}\sigma{\cal A}^{}_{\sigma\sigma '}\sigma '{\cal N}^{}_{R,\sigma '}\ ,\label{C}
\end{align}
giving  the various transmission probabilities of the junction. \cite{com2}

\subsection{The Onsager relations}

\label{ONS}
As was mentioned in Sec. \ref{intro}, there is a certain analogy between the configuration studied here and that of thermoelectric transport. In order to further pursue this point we consider the entropy production in our device, assuming that the spin imbalance in each of the two reservoirs does not vary with time and that all parts of the setup are held at the same temperature $T$. Under these circumstances  the entropy production, $\dot{S}$,  is
\begin{align}
T\dot{S}&=\sum_{\sigma}\mu^{}_{L,\sigma}I^{}_{L,\sigma}+\sum_{\sigma '}\mu^{}_{R,\sigma '}I^{}_{R,\sigma '}
\nonumber\\
&=I^{}_{L}(\mu^{}_{L}-\mu^{}_{R})+U^{}_{L}I^{\rm spin}_{L}+U^{}_{R}I^{\rm spin}_{R}\ ,\label{sd}
\end{align}
where the various currents are given in Eqs. (\ref{II}) and (\ref{IIS}).
Obviously, the first term on the right-hand side of Eq. (\ref{sd}) is the dissipation due to Joule heating. The other two terms describe the dissipation involved with  the spin currents.

As is seen from Eq. (\ref{sd}), the entropy production may be presented as a scalar product of the vector of driving forces (sometimes called ``affinities"), $\{ V\equiv (\mu_{L}-\mu_{R})/e, U_{L},U_{R}\}$  and the resulting currents, $\{eI^{}_{L},I^{\rm spin}_{L},I^{\rm spin}_{R}\}$. In the linear-response regime (see Sec. \ref{LR})
these two vectors are related to one another by a (3$\times$3) matrix ${\cal M}$, which contains the transport coefficients, 
\begin{align}
\left [\begin{array}{c}eI^{}_{L} \\I^{\rm spin}_{L} \\I^{\rm spin}_{R}\end{array}\right ]
={\cal M}\left [\begin{array}{c}V \\U^{}_{L} \\U^{}_{R}\end{array}\right ]\label{22}
\end{align}
with
\begin{align}
{\cal M}=\left [\begin{array}{ccc}
e^{2}{\cal C}^{}_{1}&e{\cal C}^{}_{2}&-e{\cal C}^{}_{3} \\    \  e{\cal C}^{}_{2}&\ \ {\cal C}^{}_{1}&-{\cal C}^{}_{4}\\   
-e{\cal C}^{}_{3}&-{\cal C}^{}_{4}&\ \ {\cal C}^{}_{1}\end{array}\right ]\ .\label{23}
\end{align}
One notes that this matrix obeys the Onsager relations: reversing the sign of the magnetic field, i.e., inverting the sign of the  Aharonov-Bohm phase $\phi$ [Eq. (\ref{phi})] and concomitantly interchanging the vibration states indices  $n$ with $n'$ and the spin indices  $\sigma$ with $\sigma'$ in Eqs. (\ref{T}) and (\ref{a}) leaves all off diagonal terms in the  matrix ${\cal M}$ unchanged.

\section{Spin-electric transport through a Rashba twister device
}


\label{SV}

\subsection{The transport coefficients}
\label{coef}

The full calculation of the transmission  matrix ${\cal A}$ that determines the transport coefficients ${\cal C}_{i}$ [see Eqs. (\ref{a}) and (\ref{C})] is quite complicated, and requires a numerical computation. 
We provide in Appendix \ref{ME} an approximate form for it, valid when the coupling of the charge carriers to the vibrational modes of the wire is weak.
The approximation is based on the different magnitudes  that coupling takes in  the magnetic Aharonov-Bohm phase and  in the Rashba one. In order to see this, it is expedient to 
present the phase factors in the transmission amplitude in the form
\begin{align}
\exp(-i\psi_{R})\exp(-i\psi_{L})\equiv e^{-i\phi}(A+i{\bf B}\cdot\sig)\ ,\label{p1n}
\end{align}
 [see Eqs. 
(\ref{vlr}), (\ref{psilr}), and (\ref{ww})]. 
Here $A$ and ${\bf B}$ are functions of the instantaneous bending angle $\theta$,  Eq. (\ref{theta}), 
\begin{align}
&A=1-2\cos^{2}(\theta)\sin^{2}(\alpha d/2)\ ,\nonumber\\
&{\bf B}=\{ 0,\cos (\theta)\sin(\alpha d),-\sin (2\theta)\sin^{2}(\alpha d/2)\}\ , \nonumber\\
&A^{2}_{}+{\bf B}\cdot{\bf B}=1\ , \label{p2n}
\end{align}
and $\phi$ is the instantaneous Aharonov-Bohm flux in dimensionless units, Eq. (\ref{phi}). 
The components of the spin-orbit vector ${\bf B}$ are given in the coordinate axes depicted in Fig. \ref{1}.

As can be observed by inserting Eq. (\ref{theta}) for  the vibration-dependent bending angle into  Eqs.  (\ref{vlr}) and (\ref{psilr}),   the effect of the electron-vibration interaction on the Rashba coupling is of  the order of the zero-point amplitude of the vibrations divided by the wire length, $a_{0}/d$. 
On the other hand, upon inserting Eq. (\ref{theta}) into Eq. (\ref{phi}) one finds that the Aharonov-Bohm phase is
\begin{align}
\phi \simeq -\frac{\pi Hd^{2}}{4\Phi^{}_{0}}\sin (2\theta^{}_{0})-\frac{\pi a^{}_{0}d H}{2\Phi^{}_{0}}\cos (\theta^{}_{0})\cos (2\theta^{}_{0})(b+b^{\dagger})\ .
\end{align}
The dynamics of the Aharonov-Bohm flux is thus determined by the flux enclosed in an area of order $a_{0}d$  divided by the flux quantum (see Appendix \ref{ME}). The latter ratio can be significantly larger than $a_{0}/d$. 
For instance, the length of a single-walled carbon nanotube is about $d=1\mu$, while the vibrations' zero-point amplitude is estimated to be $10^{-5}\mu$. This  leads to $a_{0}/d\simeq 10^{-5}$, while $(Ha_{0}d)/\Phi_{0}$ is of the order of $10^{-2}$ for magnetic fields of the order of a few Teslas (at which the effect of the magnetic field on the transport through the Rashba weak link becomes visible).

The disparity between  the way the electron-vibration coupling affects the Rashba phase factor and the manner by which it dominates the magnetic one  results in a convenient (approximate) form for the transmission matrix ${\cal A}$. \cite{com2}    We show in Appendix \ref{ME} 
that 
\begin{align}
{\cal A}=g(T,H)\left [\begin{array}{cc}{\cal A}^{}_{\rm d}&{\cal A}^{}_{\rm nd}\\ {\cal A}^{}_{\rm nd}&{\cal A}^{}_{\rm d}\end{array}\right ]\ .\label{AM}
\end{align}
Here $g$ is  the transmission of the junction in the absence of the Rashba interaction; it depends on the temperature and on the perpendicular magnetic field, 
\begin{align}
&g(T,H)=\Big (\frac{J^{}_{L}J^{}_{R}}{\epsilon^{}_{0}}\Big )^{2}\Big (\sum_{n=0}^{\infty}P(n)|\langle n|e^{-i\phi}|n\rangle |^{2}\nonumber\\
&+\sum_{n=0}^{\infty}\sum_{\ell =1}^{\infty}P(n)|\langle n|e^{i\phi}|n+\ell\rangle |^{2}\frac{2\ell\beta\omega}{e^{\ell
\beta\omega}-1}\Big )\ .\label{g}
\end{align}
This quantity is discussed extensively in Ref. \onlinecite{RIS} where one may find its detailed dependence on the temperature and on the magnetic field.
In particular, at high and low temperatures (compared to the vibration frequency)
\begin{align}
g(T,H)=\Big (\frac{J^{}_{L}J^{}_{R}}{\epsilon^{}_{0}}\Big )^{2}\Bigg \{\begin{array}{c}1-\frac{\beta\omega}{6}\frac{H^{2}}{H^{2}_{0}}\ ,\ \ \ \ \ \ \ \ \ 
\beta\omega\ll 1\ ,\\
\exp[-H^{2}_{}/H^{2}_{0}]\ ,\ \ \ \  \beta\omega\gg 1\ ,\end{array}
\end{align}
where 
$H^{}_{0}=\sqrt{2}\Phi^{}_{0}/[\pi da^{}_{0}\cos (\theta^{}_{0})\cos (2\theta^{}_{0})]$, with $a_{0}$ being the amplitude of the zero-point oscillations and $\Phi_{0}$ the flux quantum.

The spin-dependent part of the transmission is given by the matrix in Eq. 
(\ref{AM}), 
\begin{align}
{\cal A}^{}_{\rm d}+{\cal A}^{}_{\rm nd}&=1\ ,\nonumber\\
{\cal A}^{}_{\rm d}-{\cal A}^{}_{\rm nd}&=(A_{0}^{2}-B_{0}^{2})\hat{\bf n}^{}_{L}\cdot\hat{\bf n}^{}_{R}+2A^{}_{0}{\bf B}^{}_{0}\cdot\hat{\bf n}^{}_{L}\times\hat{\bf n}^{}_{R}\nonumber\\
&+2({\bf B}^{}_{0}\cdot\hat{\bf n}^{}_{L})({\bf B}^{}_{0}\cdot\hat{\bf n}^{}_{R})\ .\label{ADND}
\end{align}
Here $A_{0}$ and ${\bf B}_{0}$ are given by the values of  $A$ and ${\bf B}$ defined in Eqs. (\ref{p2n}) at equilibrium, i.e., when the angle $\theta$ there is replaced by $\theta_{0}$.
Their physical meaning  is explained in Sec. \ref{RISC}: ${\cal A}_{\rm nd}=\sin^{2}(\gamma)$, where $\gamma$ is the twisting angle of the charge carriers' spins, and ${\cal A}_{\rm d}=\cos^{2}(\gamma )$.

Using the explicit expression (\ref{AM}) for the transmission matrix ${\cal A}$
it is straightforward to find the transport coefficients ${\cal C}_{i}$. 
Retaining only terms linear in the difference between the densities of states of the spin orientations, we obtain
\begin{align}
{\cal C}^{}_{1}+{\cal C}_{4}&\simeq 8\pi g(T,H) {\cal A}^{}_{\rm d}{\cal N}^{}_{L}{\cal N}^{}_{R}\simeq {\cal C}_{2}+{\cal C}^{}_{3}\ ,\nonumber\\
{\cal C}^{}_{1}-{\cal C}_{4}&\simeq 8\pi g(T,H){\cal A}^{}_{\rm nd}{\cal N}^{}_{L}{\cal N}^{}_{R}\ ,\nonumber\\
{\cal C}^{}_{2}-{\cal C}^{}_{3}&=4\pi g(T,H){\cal A}^{}_{\rm nd}({\cal N}^{}_{L,\up}{\cal N}^{}_{R,\down}-{\cal N}^{}_{L,\down}{\cal N}^{}_{R,\up})\ ,\label{C1}
\end{align}
where ${\cal N}_{L,R}$ is the total density of states of each electronic reservoir (summed over the two spin directions).
Glancing at Eq. (\ref{II})
for the charge current and taking into account the first of Eqs. (\ref{ADND}), shows that the conductance, $G$,  of the junction is independent of the spin-orbit interaction, and is given by \cite{com5}
\begin{align}
G=4\pi e^{2}_{}{\cal N}_{L}{\cal N}^{}_{R}g(T,H)\ .
\label{GC}
\end{align}
Specific spintro-voltaic effects are considered below.

\subsection{Rashba twisting}

\label{RISC}
When the junction is not subject to a perpendicular magnetic field and the charge carriers passing through it do not collect an Aharonov-Bohm phase due to it, one may safely ignore the effect of the quantum flexural nano-vibrations of the suspended wire. Indeed, the electron-vibration coupling on the weak link is of order
$a_{0}/d\simeq 10^{-5}$  for carbon nanotubes (see the model description in Secs. \ref{GEN} and \ref{coef},  and Appendix \ref{ME}). 
This interaction  is therefore not expected to modify significantly the transmission through the wire. 

The scattering of the 
electrons' momentum, caused by the spatial constraint of their orbital motion inside the nanowire, also induces 
scattering of the electronic spins. The latter results from the spin-orbit Rashba interaction located at the wire. Consequently,  an electronic wave having  a definite  spin projection  on the magnetization vector of  the lead from which it emerges, is not a spin eigen state in the other lead.

Thus, a pure spin state $|\sigma\rangle$ in one lead becomes a mixed spin state 
in the other, 
\begin{align}
|\sigma\rangle\Rightarrow
\alpha_{1}|\sigma\rangle+\alpha_{2}|\overline{\sigma}\rangle\ ,
\label{amps}
\end{align}
with probability amplitude $\alpha_{1}$ to remain in  the original state, and probability amplitude  $\alpha_{2}$ for a spin flip ($\overline{\sigma}=-$$\sigma$).
During the propagation   through the weak link the spins
of the charge carriers are twisted, 
as is described by the transmission amplitude [see Eqs. (\ref{p1n}) and (\ref{A1})],
$A^{}_{0}+i{\bf B}^{}_{0}\cdot\sig$.  
It follows that the probability amplitude for a spin flip, $\alpha_{2}$, 
is  given by
\begin{align}
\alpha^{}_{2}=[{\cal S}^{\dagger}_{R}(A^{}_{0}+i{\bf B}^{}_{0}\cdot\sig){\cal S}^{}_{L}]^{}_{\sigma\overline{\sigma}}\ ,
\end{align}
with  ${\cal S}_{L,R}
$ given in Eq. (\ref{slr}). The Rashba  twisting angle, $\gamma$,  can now be defined by
\begin{align}
\alpha^{}_{2}=\sin(\gamma )e^{i\delta}\ ,
\end{align}
with
\begin{align}
|\alpha^{}_{2}|^{2}=\sin^{2}(\gamma)={\cal A}_{\rm nd}\ , \label{AG}
\end{align}
yielding a clear physical meaning to the transmissions ${\cal A}_{\rm d}$ and ${\cal A}_{\rm nd}$ [see Eqs. (\ref{ADND})].

The physical quantities depend only on the relative phase between $\alpha_1$ and $\alpha_2$. Therefore, we choose  $\alpha_1=\cos\gamma$.
It is then easy to check that the average of the vector ${\bf\sigma}$ in the state of Eq. (\ref{amps}) is equal to $\{\sin(2\gamma)\cos(\delta),\sin(2\gamma)\sin(\delta),\cos(2\gamma)\}$. This vector is rotated by the angle $2\gamma$ relative to its direction in the absence of the spin-orbit interaction. We call this rotation of the electronic moments in each of the two leads a ``twist" of the spins.  It is distinct from simple spin precession since the axis of this precession changes its direction during the electronic motion along the curved trajectory.

In the simplest configuration of parallel magnetizations in both electrodes, i.e., 
\begin{align}
\hat{\bf n}_{L}^{}=\hat{\bf n}^{}_{R}\equiv \hat{\bf n}\ ,
\end{align}
Eqs. (\ref{ADND}) yield
\begin{align}
\sin (\gamma )=[B_{0}^{2}-(\hat{\bf n}\cdot{\bf B}^{}_{0})^{2}]^{1/2}
\ .\label{gama}
\end{align}
Interestingly enough, in this simple configuration $\sin (\gamma)$ is determined by the component of the Rashba   vector ${\bf B}_{0}$ normal to the quantization axis of the magnetization in the electrodes.
Mechanically manipulating the bending angle that determines the direction of the Rashba vector ${\bf B}_{0}$,  one may control the twisting angle $\gamma$. Note also that had the vectors $\hat{\bf n}_{L}$ and $\hat{\bf n}_{R}$ been antiparallel to one another
then $\sin (\gamma )=[1-B_{0}^{2}+(\hat{\bf n}\cdot{\bf B}^{}_{0})^{2}]^{1/2}$.

\begin{figure}[htp]
\includegraphics[width=5cm]{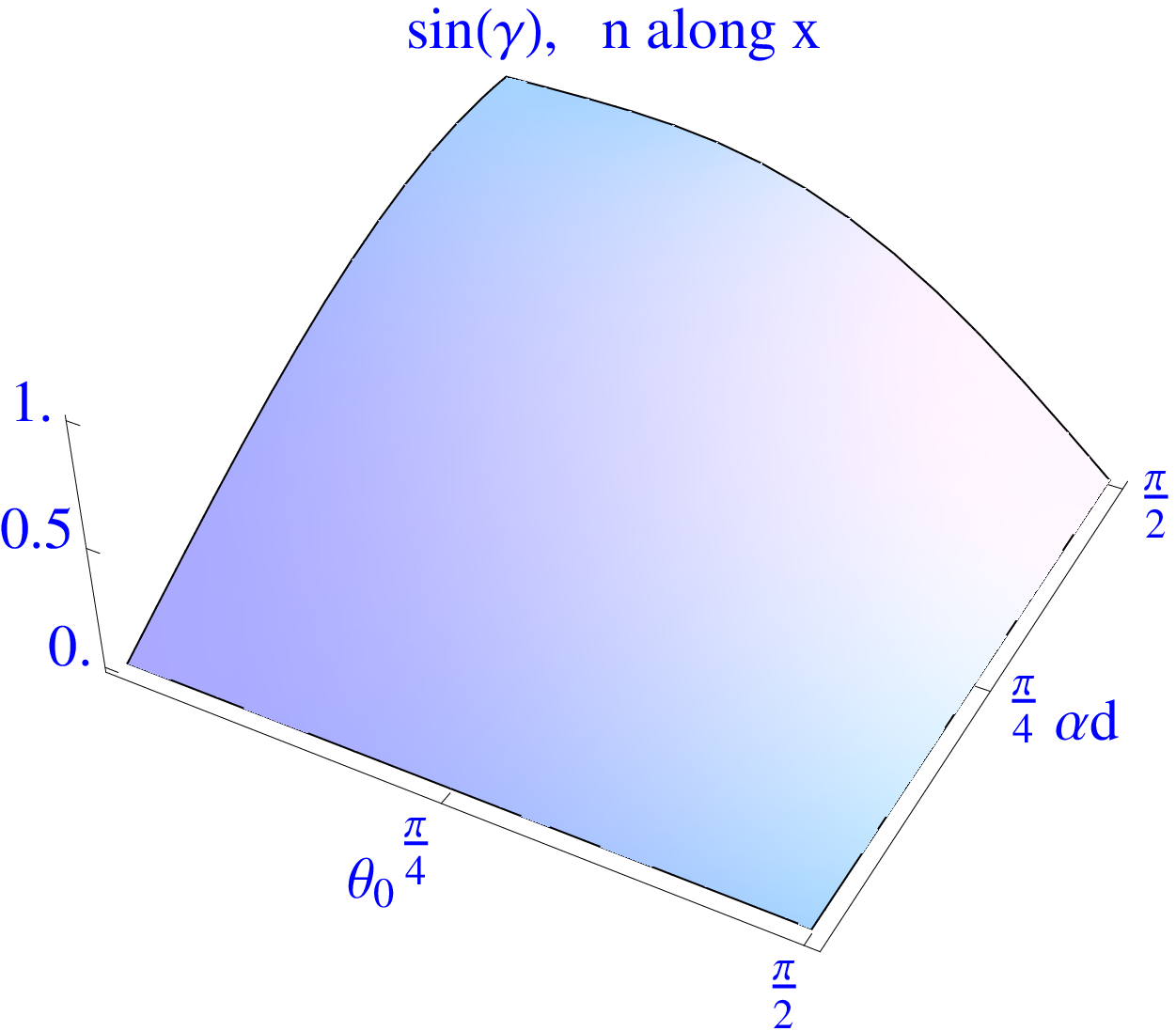}
\vspace{0.5cm}
\includegraphics[width=5cm]{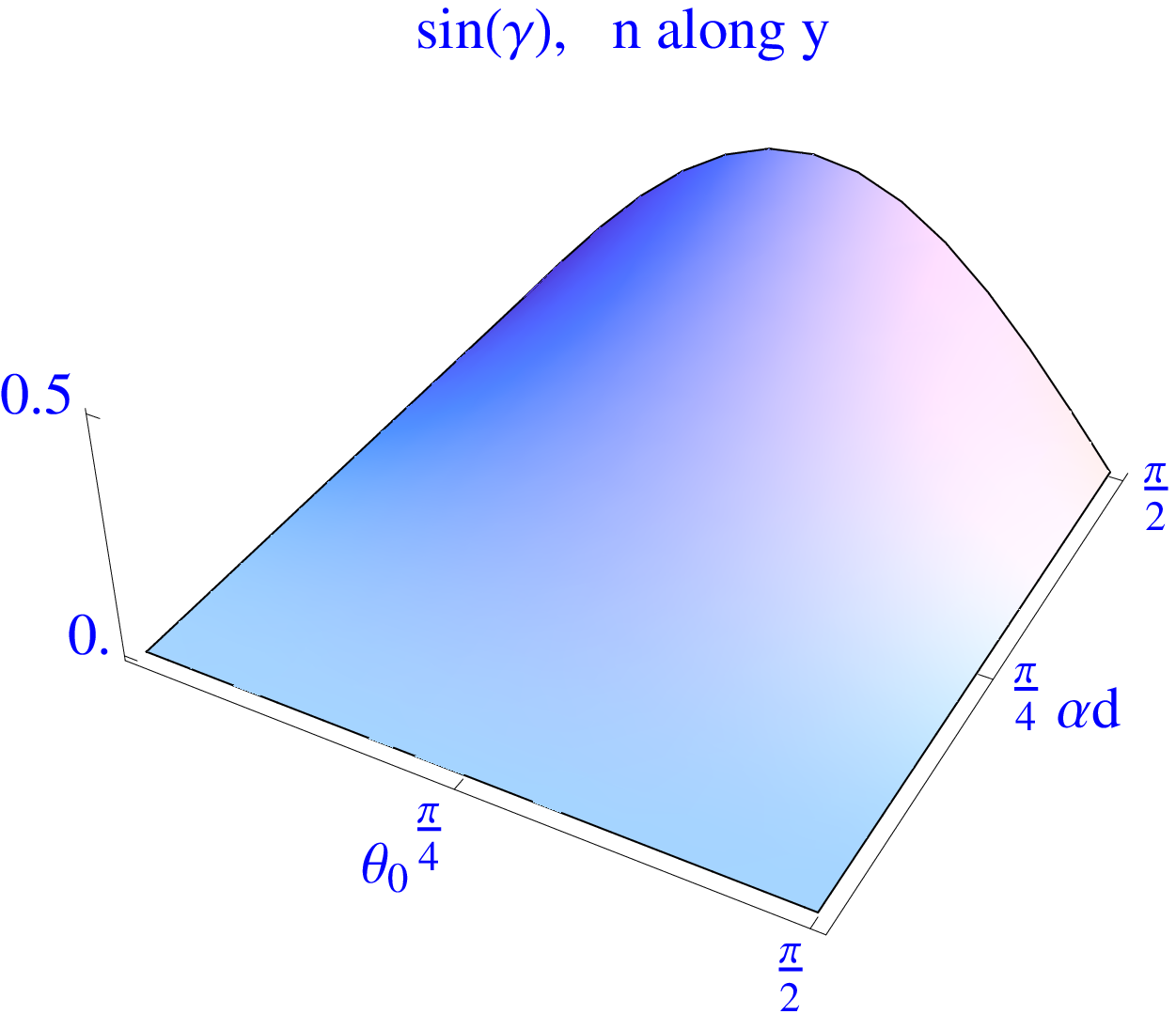}
\vspace{0.5cm}
\includegraphics[width=5cm]{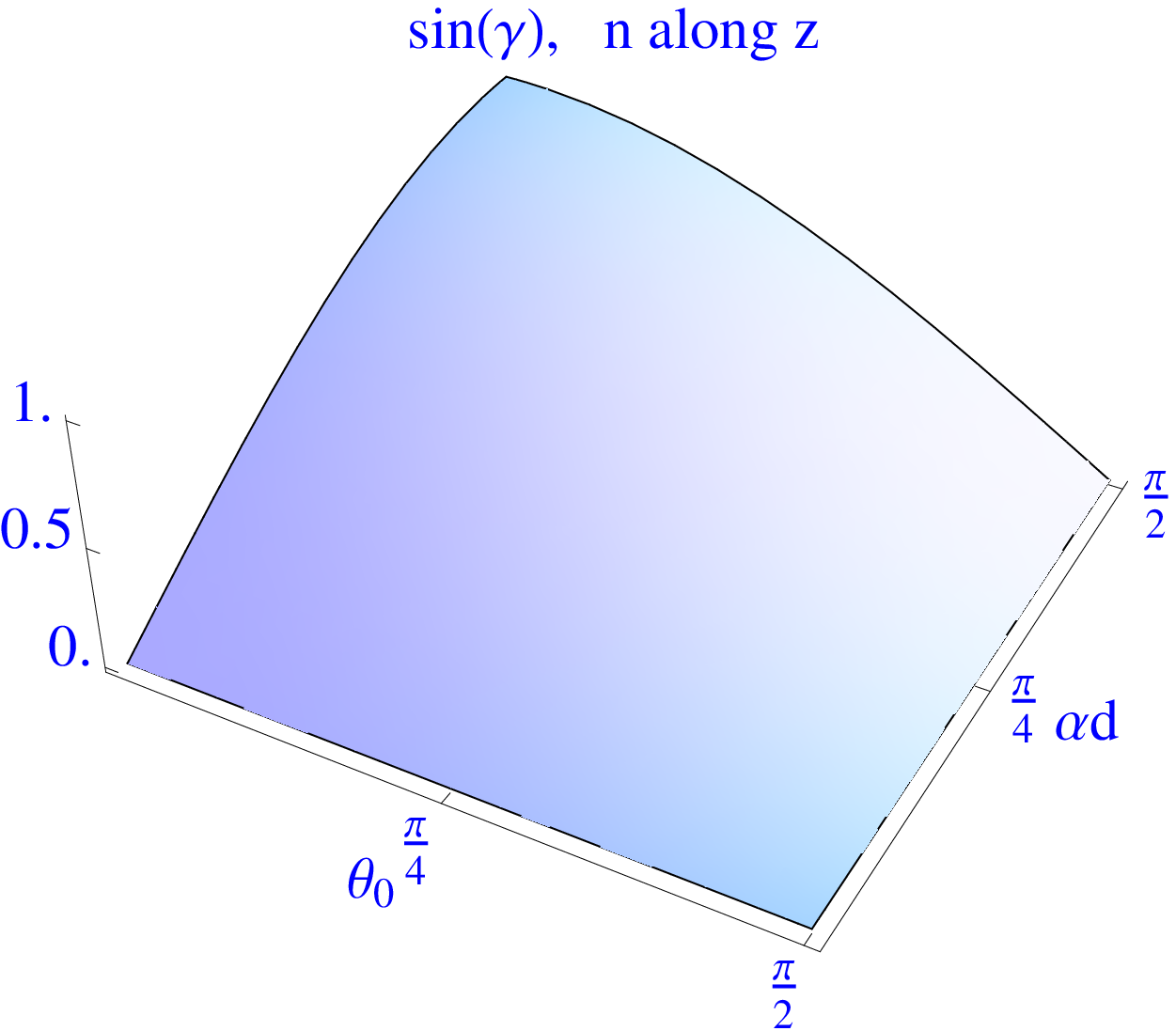}
\caption{The 
Normalized spin conductance, 
Eq. (\ref{NSC}), expressed in terms of the twisting angle 
[see Eqs. (\ref{AG}), (\ref{gama}), and (\ref{41}-\ref{43})]
as a function of  the spin-orbit coupling (scaled by the length of the junction) and on the bending angle $\theta_{0}$ of the junction (see Fig. \ref{1}). The three plots correspond to the magnetization in the leads aligned along $\hat{x}$, $\hat{y}$ and $\hat{z}$, as marked on the figures.
}
\label{3}
\end{figure}

An even more convenient way to monitor 
the twisting effect
may be realized by studying  the spintro-voltaic 
effect in an open circuit, i.e., when the total charge current vanishes. One then finds that the spin-imbalanced populations in the electrodes give rise to an electric voltage, $V_{sv}$.  Assuming  that the spin imbalances in the two reservoirs are identical, i.e., $U_{L}=U_{R}\equiv U$, Eq. (\ref{II}) yields
\begin{align}
V^{}_{sv}=\frac{{\cal C}^{}_{3}-{\cal C}^{}_{2}}{{\cal C}^{}_{1}}U\ .\label{vsv}
\end{align}
The ratio of the voltage created by the spin imbalance, $V_{sv}$, to the amount of spin imbalance in the electrodes (expressed by $U$) can be found upon using Eqs. (\ref{C1}), in conjunction with Eqs. (\ref{ADND}) and  (\ref{AG}), 
\begin{align}
V^{}_{\rm sv}=\sin^{2}(\gamma) \frac{{\cal N}^{}_{L\up}{\cal N}^{}_{R\down}-{\cal N}^{}_{L\down}{\cal N}^{}_{R\up}}{{\cal N}^{}_{L}{\cal N}^{}_{R}}U\ .
\end{align}
The voltage generated by the Rashba interaction gives directly the twisting angle; the proportionality between $V_{\rm sv}/U$ and $\sin^{2}(\gamma) $ being the magnetic mismatch parameter of the junction.

The twisting angle $\gamma$ 
determines also the  various spin conductances of the junction. From Eqs. (\ref{22}) and (\ref{23}) we find \cite{com6}
\begin{align}
&I^{\rm spin}_{L}+I^{\rm spin}_{R}=eV({\cal C}_{2}-{\cal C}^{}_{3})+(U^{}_{L}+U^{}_{R})({\cal C}_{1}^{}-{\cal C}^{}_{4})\ .
\nonumber\\
&=2\frac{G}{e^{2}}
{\cal A}^{}_{\rm nd}\Big (U^{}_{L}+U^{}_{R}+eV
\frac{{\cal N}^{}_{L,\up}{\cal N}^{}_{R,\down}-{\cal N}^{}_{L,\up}{\cal N}^{}_{R,\down}}{2{\cal N}^{}_{L}{\cal N}^{}_{R}}\Big )\ ,
\label{41}
\end{align}
where $G$ is the charge conductance, Eq. (\ref{GC}), and we have made use of Eqs. (\ref{C1}) for the ${\cal C}$'s.
One now observes that both the spin conductance, $G^{\rm spin}$ (normalized by the charge conductance)
\begin{align}
G^{\rm spin}_{}=\frac{I^{\rm spin}_{L}+I^{\rm spin}_{R}}{(U^{}_{L}+U^{}_{R})G/e^{2}}\Big |^{}_{V=0}\ ,
\label{NSC}
\end{align}
and the cross spin conductance, $G^{\rm spin}_{\times}$, (again normalized by the charge conductance),
\begin{align}
G^{\rm spin}_{\times}=\frac{I^{\rm spin}_{L}+I^{\rm spin}_{R}}{eV
G/e^{2}}\Big |^{}_{U_{L}^{}=U_{R}^{}=0}\ ,
\label{43}
\end{align}
are determined by ${\cal A}_{\rm nd}$, that is by the twisting angle $\gamma$, Eq.
(\ref{AG})
(the second requires the asymmetry  in the spin-resolved densities of states).

\begin{figure}[htp]
\includegraphics[width=5cm]{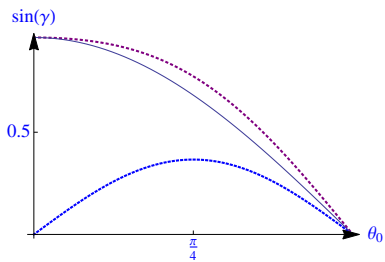}
\caption{
The 
Normalized spin conductance, 
Eq. (\ref{NSC}), expressed in terms of the twisting angle 
[see Eqs. (\ref{AG}), (\ref{gama}), and (\ref{41}-\ref{43})
as a function of 
the wire $\theta_{0}$, when the magnetizations in the leads are along $\hat{x}$, (upper dotted line) along $\hat{y}$ (lower dotted line), and along $\hat{z}$ (full curve). Here $\alpha d=1.3$ radians.
}
\label{4}
\end{figure}

For parallel magnetizations in the leads, the twisting angle [see Eq. (\ref{gama})]
depends solely on the spin-orbit coupling and on the equilibrium value of the bending angle. We plot in Fig. \ref{3} the 
dependence of $\sin(\gamma )$ on these two factors; for convenience, 
we display in Fig \ref{4} a cut of these plots. Perhaps the most significant features of these plots is the spin-twisting angle at two special values of the bending angle. Firstly we note the disappearance of the spin twisting for any direction of the polarizations in the leads at $\theta_{0}=\pi/2$. This can be easily understood within  a classical picture for the spin rotation caused by the Rashba interaction. 
The spin evolution of the tunneling electron can be regarded as a rotation around an axis given by the vectorial product of the velocity and the electric field (directed along $\hat{z}$ in our configuration). At
this value of $\theta_{0}$ the tunneling trajectory is oriented along the $\hat{y}-$axis (because then $x_{R}=x_{L}=0$)       
and so the electron ``rattles" back and forth along $\hat{y}$. This leads to a cancellation of the Rashba contribution to the tunneling phase [see Eq. (\ref{psilr})]. 
The other special case is when the wire is not bended, i.e., $\theta_{0}=0$. The spin twisting for leads' magnetizations along $\hat{y}$ vanishes,  while for devices with ferromagnetic magnetizations along the  $\hat{x}-$ or $\hat{z}-$directions it reaches its maximal value, $\sin(\alpha d)$. The reason for this has also to do with the orientation of the spin rotation-axis. At small values of $\theta_{0}$ the electronic 
trajectory is primarily along $\hat{x}$. Then, when the spin of the incident electron is directed along $\hat{y}$ (as is the case described by the dotted lower curve in Fig. \ref{4}) it is parallel to the rotation axis and no rotation is taking place. As opposed, when the spin of the incident electron is oriented along $\hat{x}$ or $\hat{z}$, it is perpendicular to the rotation axis, leading to a full rotation.
For carbon nanotubes,  \cite{SONT}  for which the energy gap induced by the spin-orbit coupling is 0.37 meV, the strength $\alpha $ is about 10$^{4}$ cm$^{-1}$, and therefore tubes of length of a  few microns are expected to produce the maximal twisting.

\section{Conclusions}

\label{CON}

In this paper we have shown that one can have additional spintro-electric functionalities if one uses a vibrating  suspended weak link, with both a magnetic flux and an (electric field dependent) Rashba spin-orbit interaction. The twisting of the electronic spins as they move between the (spin polarized) electrodes can be manipulated by the bias voltage, the bending of the weak link wire and the polarisations of the spins in the electrodes.

The traditional picture of twisting of the electronic spin is viewed quantum-mechanically as a splitting of the electronic wave in  spin space. We have shown that the twisting angle, which determines the probability amplitude of such a splitting can be measured electrically through a spintro-voltaic effect. The Rashba device proposed in this paper is therefore a promising component to be incorporated into mesoscopic electronic circuits where quantum coherence determines various interference effects of electronic waves, split
in both momentum - and spin spaces.



\begin{acknowledgments}
This work was supported by the Israeli Science Foundation (ISF) and the US-Israel Binational Science Foundation (BSF). Financial support from Swedish VR is greatly acknowledged.
RIS acknowledges the hospitality of Ben Gurion University and the support of the Distinguished Scientist Visitor Program. \end{acknowledgments}

\appendix

\section{The tunneling Hamiltonian}

\label{TUNHAM}

Here we add details of the derivation of the effective tunneling Hamiltonian, Eqs. (\ref{HTUN}) and (\ref{W}).

The Hamiltonian of the system comprises the Hamiltonian of the two leads, modelled for simplicity by free electron gases [see the discussion
at the beginning of Sec. \ref{GEN}, culminating with Eqs. (\ref{Fermi})], the free vibration Hamiltonian, ${\cal H}_{\rm vib}=\omega b^{\dagger}b$
[see the discussion around Eq. (\ref{theta})], the Hamiltonian of the dot  representing the wire, ${\cal H}_{\rm dot}=\epsilon_{0}\sum_{\sigma}c^ {\dagger}_{0\sigma}c^{}_{0\sigma}$, and the tunneling Hamiltonian, 
\begin{align}
{\cal H}^{}_{\rm tun}&=\sum_{k,\sigma,\sigma '}(V^{}_{k\sigma \sigma '}c^{\dagger}_{0\sigma}c^{}_{k\sigma '}
+{\rm H.c. })\nonumber\\
&+\sum_{p,\sigma,\sigma '}(V^{}_{p\sigma \sigma '}c^{\dagger}_{p\sigma}c^{}_{0\sigma '}+{\rm H.c. })\ .
\label{ht}
\end{align}
The electron gas states in the left (right) lead are indexed by $k$ ($p$) and have energies $\epsilon_{k}$ ($\epsilon_{p}$). We denote
by $c_{k\sigma}$ ($c_{p\sigma}$) the annihilation operators for the leads, and  by  $c_{0\sigma}$ that for the localized level representing the wire [see Fig. \ref{2})].
The tunneling amplitudes in Eq. (\ref{ht}) are given in Eqs. (\ref{vlr}) and (\ref{psilr}).
We consider a non-resonant case, where the localized level
is far above the energies of the occupied states in both leads  (i.e., no energy level on the wire is close enough to $\epsilon^{}_0$  
to be 
involved in  inelastic tunneling via a real state). This allows us to exploit the tunneling as an expansion parameter \cite{RIS} and to
preform a unitary transformation
which replaces the wire by an effective direct tunneling between the leads through virtual states
\begin{align}
{\cal H}^{\rm e}_{\rm tun}=
\sum_{k,p}(c^{\dagger}_{k}W^{\dagger}_{kp}c^{}_{p}+{\rm H.c.})\ ,
\end{align}
with  $W$  given  in Eq. (\ref{W}) (using matrix notations in spin space).

\section{The transmission matrix 
}

\label{ME}

Here we detail the approximate calculation of the transmission probability ${\cal A}$, Eq. (\ref{a}), confining ourselves to the case of weak coupling of the electrons to the vibrational modes.

We begin by re-writing the transmission ${\cal T}$, Eq. (\ref{T}), in the form
\begin{align}
\label{A1}
{\cal T}^{}_{nn'\sigma\sigma '}&=\Big (\frac{J^{}_{L}J^{}_{R}}{\epsilon^{}_{0}}\Big )^{2}|\langle n|e^{-i(\phi^{}_{0}+\Delta\phi)}\\
&\times\Big [{\cal S}^{\dagger}_{R}[A^{}_{0}+\Delta A+i({\bf B}^{}_{0}+\Delta {\bf B})\cdot\sig]{\cal S}^{}_{L}\Big ]^{}_{\sigma' \sigma }|n'\rangle |^{2}\ .\nonumber
\end{align}
Here we have used Eqs. (\ref{p1n}) and (\ref{p2n}), expressing $\phi$,  $A$, and ${\bf B}$ as the sums of their equilibrium values, $\phi_{0}$, $A_{0}$, and ${\bf B}_{0}$  [i.e., with $\theta$ in Eqs.  (\ref{phi}) and (\ref{p2n}) replaced by $\theta_{0}$] and  their dynamical parts that include the vibrations' operators and are denoted by $\Delta\phi$, $\Delta A$, and $\Delta{\bf B}$. 
It is straightforward to verify that the terms including $\Delta A+i\Delta{\bf B}\cdot\sig$
contribute only when the electron-vibration interaction is accounted for at least to second order. In view of the smallness of the effect  that interaction has on the Rashba coupling, as opposed to its effect on the magnetic phase (see the discussion in Sec. \ref{coef})  we omit those terms, keeping the electron-vibration interaction only in the magnetic phase.
As a result we obtain
\begin{align}
{\cal T}^{}_{nn'\sigma\sigma '}&\simeq  \Big (\frac{J^{}_{L}J^{}_{R}}{\epsilon^{}_{0}}\Big )^{2}
\langle n|e^{-i\phi}|n'\rangle|^{2}|{\cal R}
^{}_{\sigma' \sigma }|^{2}\ ,
\end{align}
with  the spin-dependent part of ${\cal T}$ given by
\begin{align}
{\cal R}={\cal S}^{\dagger}_{R}(A^{}_{0}+i{\bf B}^{}_{0}\cdot\sig ){\cal S}^{}_{L}\ .
\end{align}
One  notes that
\begin{align}
{\rm Tr}\{{\cal R}{\cal R}^{\dagger}_{}\}=2\ ,
\end{align}
and 
\begin{align}
&{\rm Tr}\{{\cal R}\sigma^{}_{z}{\cal R}^{\dagger}_{}\sigma^{}_{z}\}=2[(A_{0}^{2}-B_{0}^{2})\hat{\bf n}^{}_{L}\cdot\hat{\bf n}^{}_{R}\nonumber\\
&+2A_{0}^{}{\bf B}^{}_{0}\cdot\hat{\bf n}^{}_{L}\times\hat{\bf n}^{}_{R}
+2({\bf B}^{}_{0}\cdot\hat{\bf n}^{}_{L})({\bf B}^{}_{0}\cdot\hat{\bf n}^{}_{R})]\ ,
\end{align}
where we have used
\begin{align}
{\cal S}^{}_{L,R}\sigma^{}_{z}{\cal S}^{\dagger}_{L,R}=\hat{\bf n}^{}_{L,R}\cdot\sig\ .
\end{align}

The matrix ${\cal R}$ can be written as
\begin{align}
{\cal R}=C+i{\bf D}\cdot\sig, 
\end{align}
with $C^{2}_{}+D^{2}_{x}+D^{2}_{y}+D^{2}_{z}=1$. Therefore,  the diagonal elements of the matrix ${\cal R}$ are two complex conjugate numbers, and so are also the off diagonal 
elements.  This implies that the diagonal elements of $|{\cal R}_{\sigma '\sigma}|^{2}$ are equal to one another, and the off diagonal ones are also identical. To derive explicit expressions for them, we note that the matrix ${\cal R}$ has the  property
\begin{align}
&{\rm Tr}\{C+i{\bf D}\cdot\sig)\sigma^{}_{z}(C-i{\bf D}\cdot\sig )\sigma^{}_{z}\}\nonumber\\
&=2(C^{2}+D^{2}_{z}-D^{2}_{x}-D^{2}_{y})\ .
\end{align}
It follows that
the diagonal  matrix elements (in spin space) of the transmission  are ${\cal T}_{nn',\sigma\sigma }\equiv {\cal T}^{}_{nn',{\rm d}}\equiv
C^{2}+D^{2}_{z}$   and 
the off diagonal ones are ${\cal T}_{nn',\sigma\overline{\sigma} }\equiv {\cal T}^{}_{nn',{\rm nd}}\equiv
D^{2}_{x}+D^{2}_{y}$,  
with
\begin{widetext}
\begin{align}
&{\cal T}^{}_{nn',{\rm d}}
=\frac{1}{2}\Big (\frac{J^{}_{L}J^{}_{R}}{\epsilon^{}_{0}}\Big )^{2}|\langle n|e^{-i\phi}|n'\rangle|^{2}\Big [1+\Big ((A_{0}^{2}-B_{0}^{2})\hat{\bf n}^{}_{L}\cdot\hat{\bf n}^{}_{R}+2A^{}_{0}{\bf B}^{}_{0}\cdot\hat{\bf n}^{}_{L}\times\hat{\bf n}^{}_{R}+2({\bf B}^{}_{0}\cdot\hat{\bf n}^{}_{L})({\bf B}^{}_{0}\cdot\hat{\bf n}^{}_{R})\Big )\Big ]\ ,\nonumber\\
&{\cal T}^{}_{nn',{\rm nd}}
=\frac{1}{2}\Big (\frac{J^{}_{L}J^{}_{R}}{\epsilon^{}_{0}}\Big )^{2}|\langle n|e^{-i\phi}|n'\rangle|^{2}\Big [1-\Big ((A_{0}^{2}-B_{0}^{2})\hat{\bf n}^{}_{L}\cdot\hat{\bf n}^{}_{R}+2A^{}_{0}{\bf B}^{}_{0}\cdot\hat{\bf n}^{}_{L}\times\hat{\bf n}^{}_{R}+2({\bf B}^{}_{0}\cdot\hat{\bf n}^{}_{L})({\bf B}^{}_{0}\cdot\hat{\bf n}^{}_{R})\Big )\Big ]\ .\label{ddnd}
\end{align}
\end{widetext}

Returning now to the transmission matrix ${\cal A}$, Eq. (\ref{a}), we find that it can be factorized into a temperature and magnetic filed dependent factor, and a spin-dependent factor, so that it takes the form given in the main text, Eqs. (\ref{AM}), (\ref{g}), and (\ref{ADND}).


\begin{thebibliography}{999}
\bibitem{com} 
Note that the situation at hand is more complicated than a simple spin rotation. While moving along a particular spatial path, the electron spin  rotates around an axis  whose orientation is constantly changing.
\bibitem{com3} 
This means that  the internal relaxation taking place within the electrodes is much faster as compared with the charge relaxation in the weak link. This very fact allows us  to assume local equilibrium in each of the reservoirs.
\bibitem{com1}
The case of nonmagnetic electrodes poses no problem, as the orientation of the spins there can always be chosen as that of the injected electrons.
\bibitem{com2m}

We expect this dependence to be weak
when the spin biases $U_{L,R}$  are much smaller than the Curie temperature in the magnetic leads
\bibitem{com2}
Note  the peculiar way of defining the transmission as given  in Eq. (\ref{T}): it has dimensions of energy squared because it is not yet multiplied by the appropriate densities of states. The (dimensionless) transmission probabilities are given in Eqs. (\ref{C}).
\bibitem{com5}
It should be noted that had we included the electron-vibration interaction to higher orders, we would have obtained a tiny modification of the conductance due to the spin-orbit interaction.
\bibitem{com6}
Because of charge conservation one has  $\sum_{\sigma}I^{\rm spin}_{\sigma}=2I^{\rm spin}_{\up}=-2I^{\rm spin}_{\down}$.
\end{thebibliography}
\end{document}